\documentclass[aps,superscriptaddress,twocolumn,showpacs,floatfix]{revtex4}
\usepackage{color,graphicx}
\usepackage{nicefrac}
\usepackage{float}
\usepackage{here}
\usepackage{array}
\usepackage{booktabs}
\usepackage{url}
\usepackage{multirow}
\usepackage{amsmath, amsthm, amssymb}
\usepackage[T1]{fontenc}

\begin{document}

\title{The impact of social integration on student persistence \\ in introductory Modeling Instruction courses}

\pacs{01.40.Fk, 01.40.gb, 01.40.Ha}

\keywords{social network analysis, centrality, retention, persistence}

\author{Justyna P. Zwolak}
\affiliation{STEM Transformation Institute, Florida International University, Miami, Florida 33199}

\author{Eric Brewe}
\affiliation{STEM Transformation Institute, Florida International University, Miami, Florida 33199}
\affiliation{Department of Teaching and Learning, Florida International University, Miami, Florida 33199}
\affiliation{Department of Physics, Florida International University, Miami, Florida 33199}

\begin{abstract} 
Increasing student retention and persistence -- in particular classes or in their major area of study -- is a challenge for universities. Students' academic and social integration into an institution seems to be vital for student retention, yet, research on the effect of interpersonal interactions is rare. Social network analysis is an approach that can be used to identify patterns of interaction that contribute to integration into the university. We analyze how students position within a social network in a Modeling Instruction (MI) course that strongly emphasizes interactive learning impacts their persistence in taking a subsequent MI course. We find that students with higher centrality at the end of the first semester of MI are more likely to enroll in a second semester of MI. While the correlation with increased persistence is an ongoing study, these findings suggest that student social integration influences persistence.
\end{abstract}

\maketitle

\section{Introduction}

The publication of Vincent Tinto's student integration model marks the start of the dialogue on undergraduate retention.  In his model, Tinto introduced the notion of ``external'' (e.g., families, neighborhoods, work settings) and ``internal'' (e.g., learning groups within a classroom, residence halls) communities that affect student integration into the social and academic environment of the university \cite{Tinto75-DHE,Tinto97-CAC}. Increasing the retention of students in a particular course and their persistence in continuing through their major area of study and finishing their degree is a big challenge for universities. Based on the work of Tinto and others increasing social and academic integrations is one of prime targets to increase persistence. 

Social network analysis (SNA) is a well-suited approach to study student academic and social integration. SNA can be used to identify patterns of interaction that contribute to integration into the university. It provides a methodology to assess the effect of interpersonal interactions on students' persistence. While students' academic and social integration into an institution seems to be essential to student retention, effective implementation of measures to prevent losing students is sparse. Developing network methodologies for studying retention and persistence among university students is a newly-forming research area \cite{Brewe12-SNA,Forsman14-CSN}.

We use SNA techniques to address questions of retention and persistence of students at Florida International University (FIU) -- a large, Hispanic Serving Institution. In particular, we analyze how student's position within a social network in an introductory mechanics Modeling Instruction (M-MI) course impacts their persistence in taking a subsequent electricity and magnetism MI (EM-MI) course. Modeling Instruction is a guided-inquiry interactive-engagement method of teaching that organizes instruction around building, testing and applying a handful of scientific models that represent the content core of physics. Instead of relying on lectures and textbooks, the MI program emphasizes active student construction of conceptual and mathematical models in an strongly interactive learning community. It is therefore an important case for studying the effects of building student communities on promoting persistence. 

\section{Methodology}\label{sec:method}

\begin{figure}[t]
 \includegraphics[width=0.45\textwidth]{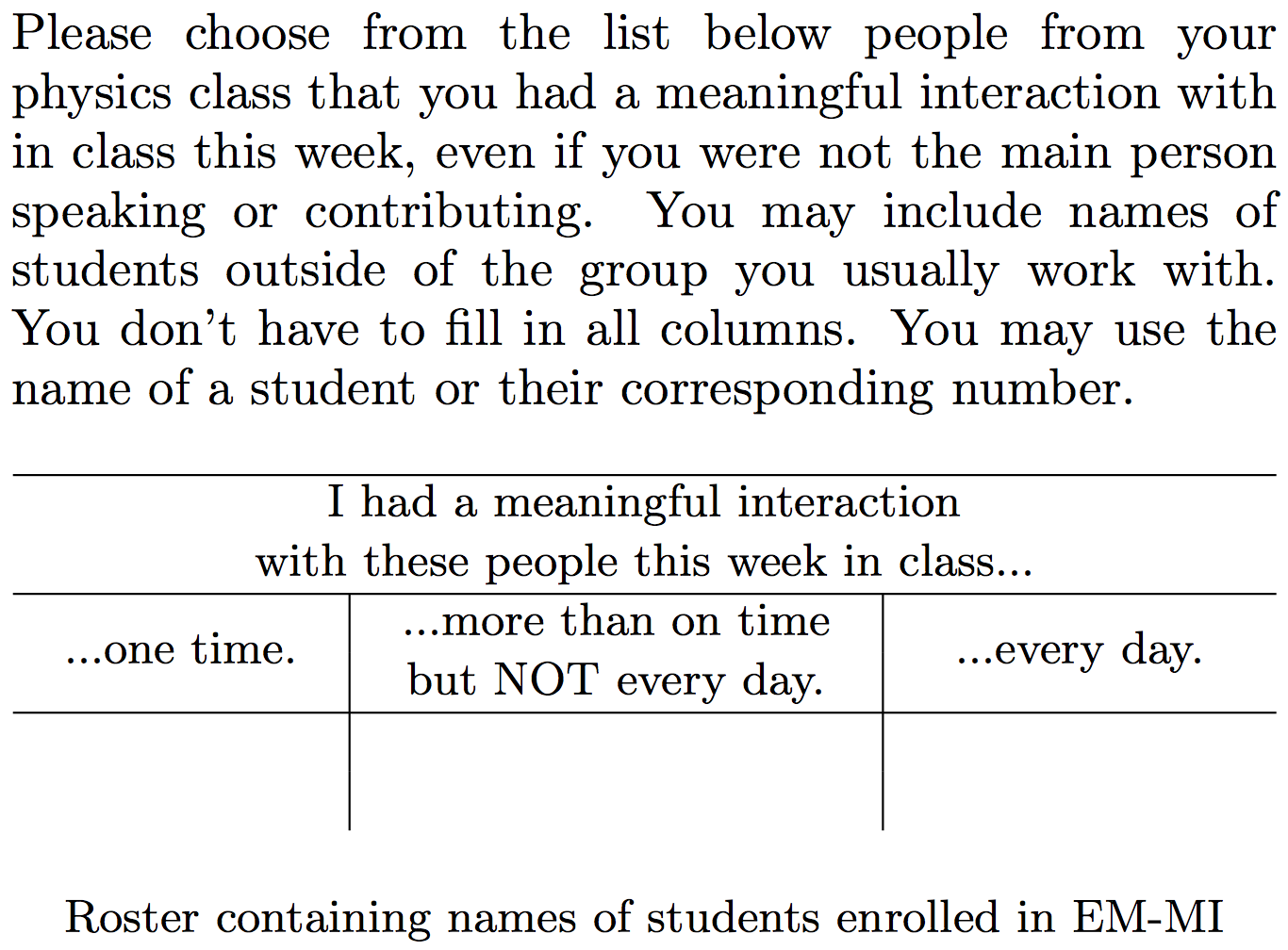}
 \caption{An excerpt of the SNA survey presented to students in a Spring 2015 semester.}
\label{fig:SNAsp15}
\end{figure}

To collect social network data we have developed a pencil and paper survey that was administered in the introductory mechanics MI course in the Fall 2014. Every four weeks throughout the semester students were asked the following question:
\begin{quote}
Name the individual(s) (first and last name) you had a meaningful classroom interaction* with today, even if you were not the main person speaking or contributing. \emph{(You may include names of students outside of the group you usually work with)} \\
\small

*A classroom interaction includes but is not limited to people you worked with to solve physics problems and people that you watched or listened to while solving physics problems.
\normalsize
\end{quote}
In the Spring semester, we presented students in the electricity and magnetism MI course with a modified version of the survey, containing a roster with names of all students enrolled in the course and a weighted version of the question about interactions, as shown in Fig.~\ref{fig:SNAsp15} .

The SNA data were collected over one semester of M-MI course (Fall 2014) and one semester of EM-MI course (Spring 2015). Both sections were taught by the same instructor accompanied by teaching assistants (the same two TAs in both semesters) and learning assistants (three LAs in a Fall semester and two LAs in a Spring semester, one overlapping person). In each semester we collected SNA data five times throughout the duration of the course. The total number of students enrolled in the M-MI was $73$ and it was $74$ for the electricity and magnetism MI. Both MI courses were taken by $40$ students and a second semester of physics in a more traditional arrangement was taken by $10$ students from M-MI. The response rates on all surveys but one were over $75\%$ and therefore we disregarded the survey with an unusually low return ($43\%$) from the analysis (the last survey in the Fall semester). In our analysis we are using the last valid survey from the Fall semester, that is SNA4. 

SNA uses the notion of nodes (in our case students enrolled in M-MI) and edges (the interactions identified by students in the survey) to represent the network. From a graph theoretic perspective, the relative importance of a node within a graph is determined using \emph{centrality measures}. To answer a question: ``Who are the most important nodes in a network?'' one has to determine how \emph{central} each node is \cite{Note1}. Evaluating the relative position of nodes in the network helps to understand the network and their participants.

\begin{figure}[b]
 \includegraphics[width=0.45\textwidth]{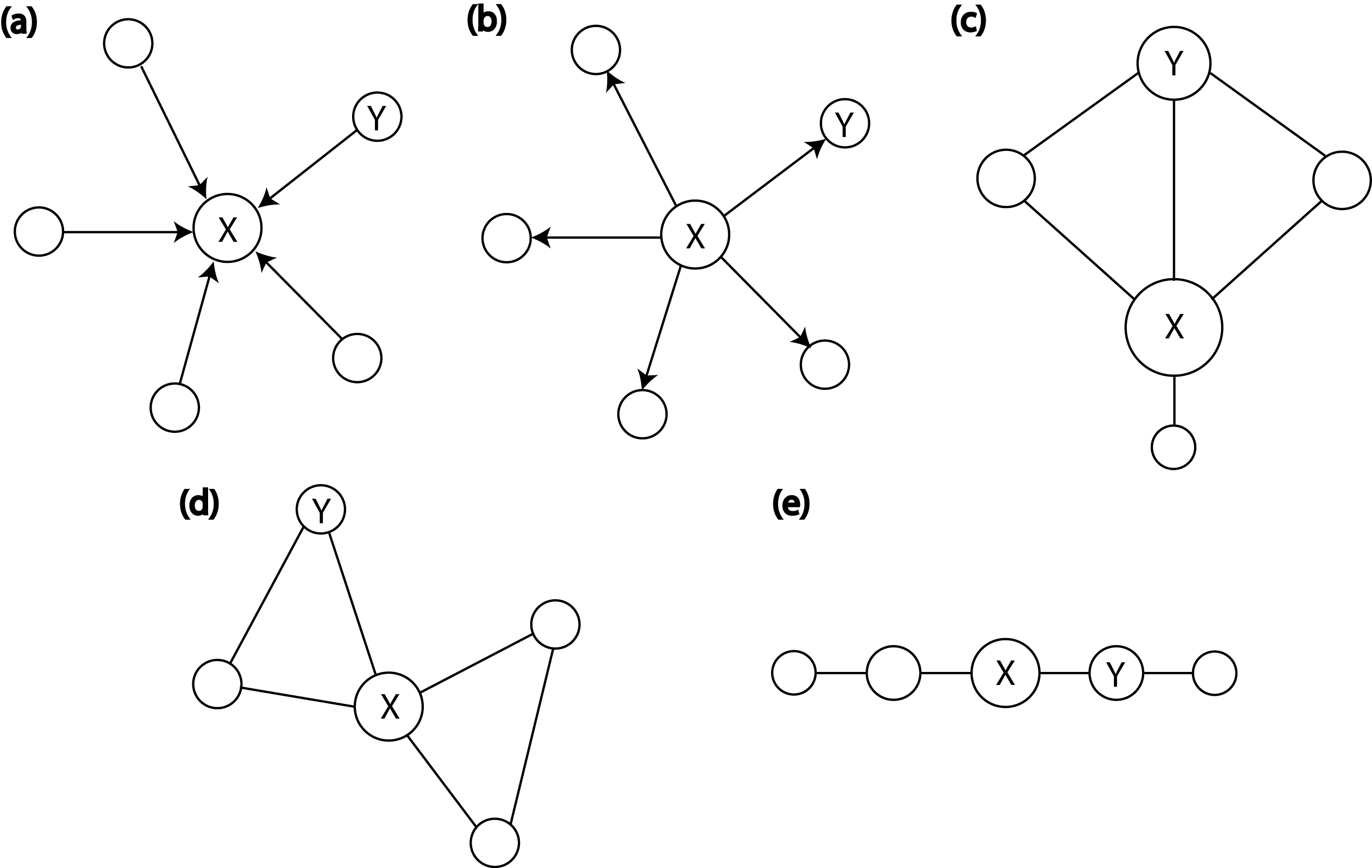}
 \caption{In each of the following networks, $X$ has higher centrality than $Y$ according to
(a) indegree, (b) outdegree, (c) eigenvector, (d) betweenness and (e) closeness.}
\label{fig:centralities}
\end{figure}

There are various measures of centrality that quantify the importance of nodes and edges. In this paper we will focus on the four most commonly used measures: degree, eigenvector, betweenness and closeness (see Fig.~\ref{fig:centralities}).  

The \emph{degree} centrality of a node $i$, $C_D(i)$, is the number of edges connected to it,
\small
\begin{equation*}
C_D(i)=\sum_{j=1}^n x_{ij}=\sum_{j=1}^n x_{ji},\\
\end{equation*}
\normalsize
where $x_{ij}$ is the value of the edge from node $i$ to node $j$ (the value being either $1$ if the tie is present or $0$ otherwise) and $n$ is the number of nodes in a network. In the case of a directed network, that is a network that takes into an account the origin of an edge, one can define two additional measures of degree centrality: \emph{indegree} (the number of ties directed to the node, can be interpreted as popularity) and \emph{outdegree} (the number of ties that the node directs to others, can be interpreted as sociability):
\small
\begin{equation*}
C_{inD}(i)=\sum_{j=1}^n x_{ji} \hspace{0.5in} C_{outD}(i)=\sum_{j=1}^n x_{ij}.
\end{equation*}
\normalsize

The \emph{eigenvector} centrality is the sum of a node's connections to other nodes weighted by their degrees and it measures the influence of a node in a network. It is given by an eigenvector, $\vec{\boldsymbol{C}}_E$, of an adjacency matrix, $\boldsymbol{A}$, corresponding to the greatest eigenvalue, $\lambda_{max}$, that is
\small
\begin{equation*}
\boldsymbol{A}^T\vec{\boldsymbol{C}}_E=\lambda_{max}\vec{\boldsymbol{C}}_E.\\
\end{equation*}
\normalsize
$\boldsymbol{A}$ is a matrix related to a graph by $a_{ij} = 1$ if a node $i$ is connected to a node $j$ by an edge and $0$ if it is not and $\vec{\boldsymbol{C}}_E$ is a vector containing the centralities of all nodes in the network.

The (in/out)degree and eigenvector centralities are very intuitive and relatively easy to calculate. However, they are all local measures and the network outside of the immediate vicinity of a node -- i.e., outside the ``ego network'' -- has no influence on them.

The \emph{betweenness} quantifies the number of times a node acts as a bridge along the shortest path linking two other nodes. It captures the importance of a position within a whole network and can be interpreted as a measure of how much control over the flow of information a node has. It's given by 
\small
\begin{equation*}
C_B(k)=\sum_{i\neq j \neq k}^n \frac{\sigma_{ij}(k)}{\sigma{ij}}
\end{equation*}
\normalsize
where $\sigma_{ij}(k)$ is the number of shortest paths linking node $i$ to node $j$ that pass through node $k$, $\sigma_{ij}$ the number of shortest paths linking node $i$ to node $j$.

\begin{figure}[t]
 \includegraphics[width=0.45\textwidth]{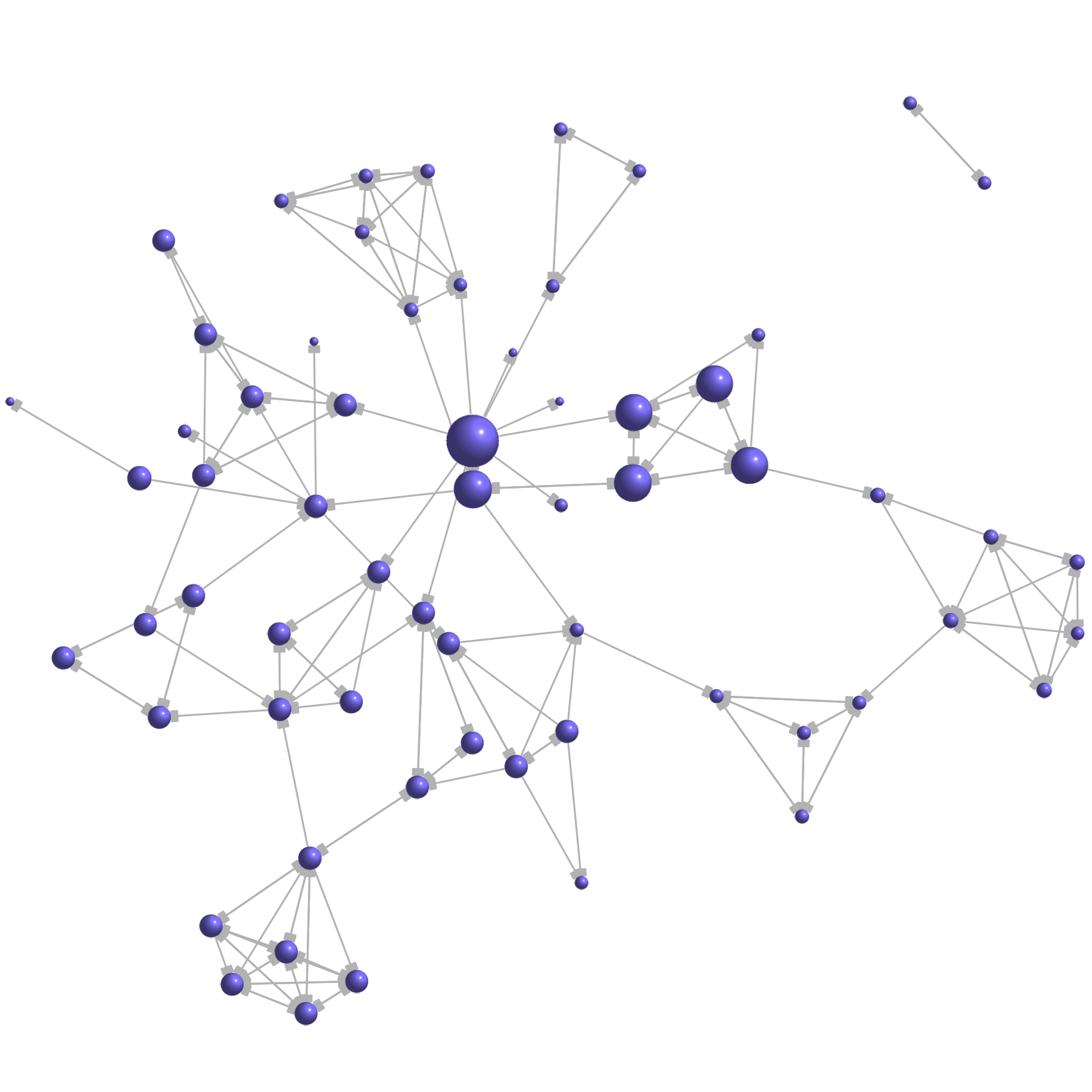}
 \caption{(Color online) The graph representation of the social network resulted from the SNA4 data. The nodes represent students enrolled in the M-MI in Fall 2014 and ties represent the directed interactions. The size of each node corresponds to student's closeness centrality.}
\label{fig:SNA4-network}
\end{figure}

The \emph{closeness} is the inverse of the sum of distances from all other nodes. It emphasizes a node's independence -- a node that is close to many other nodes can easily reach others without having to rely much on intermediaries, thus gaining an easy access to information in the network. It is a measure of how near an individual is to all other nodes in a network. Closeness is defined as
\small
\begin{equation*}
C_C(i)=\Big[\sum_{j=1}^n d_{ij}\Big]^{-1}
\end{equation*}
\normalsize
where $d_{ij}$ is the shortest distance connecting node $i$ to node $j$.  The network from survey SNA4 using the closeness as a measure of importance is visualized in Fig.~\ref{fig:SNA4-network}.

To investigate correlations between the students' centralities, gender, ethnicity, major of study, final grade and their persistence in MI, a logistic regression model (LRM) was used. To avoid confounding factors we performed multivariate logistics regression. All variables significant for the univariate analysis were incorporated into the multivariate model. The comparison of goodness of fit of multivariate and univariate models was performed using the likelihood ratio test, with the null hypothesis stating that the univariate model is a better predictor of the persistence. The  variance inflation factor (VIF) was calculated to estimate how much the variance of a coefficient was inflated because of linear dependence with other predictors. Finally, the mutual information approach was used to find the most significant split into the predicting/non-predicting categories for each of the centrality measures and the chi-square test was used to verify significance of this split  \cite{Note3}. For the statistical analysis we used the R program \cite{R}. We considered results with $p < 0.05$ as significant. 

\section{Findings}

We analyze how a student's position within a social network in a M-MI course, which strongly emphasizes interactive learning, impacts their persistence in taking a subsequent EM-MI course. We consider two cases: (1) students persistence in physics, i.e., taking any form of the second semester physics, and (2) students persistence in MI. We are interested in interactions between students and therefore we excluded from the network all instructional staff. Using the Wilcoxon rank-sum test we found no evidence for a statistically significant differences between the two population medians (i.e., with and without instructors) for all centralities but closeness. Thus, for this last measure we considered two cases -- without (closeness) and with (closenessINS) instructors.

\begin{table}[b]
\renewcommand{\arraystretch}{1.1}
\renewcommand{\tabcolsep}{6pt}
\caption{Summary of the univariate logistic regression for persistence as predicted by various centrality measures. We considered only networks without instructional staff for all measures except closeness, for which we analyzed two cases. ClosenessINS indicates a network with instructors. Significant p-values are marked with an asterisk.}
\begin{tabular}{p{1.8cm}p{1.1cm}p{1.1cm}p{1.1cm}p{1.1cm}} \hline
 & \multicolumn{2}{c}{In physics} & \multicolumn{2}{c}{In MI} \\
Centrality & Estimate & p-value & Estimate & p-value \\ \hline
Total degree & \hspace{4mm}$0.20$ & $0.016^{*}$ & \hspace{4mm}$0.12$ & $0.106$  \\ 
In degree & \hspace{4mm}$0.37$ & $0.018^{*}$ & \hspace{4mm}$0.22$ & $0.135$ \\ 
Out degree  & \hspace{4mm}$0.27$ & $0.061$ & \hspace{4mm}$0.16$ & $0.188$\\
Eigenvector & \hspace{4mm}$0.76$ & $ 0.516$ & \hspace{4mm}$0.38$ & $0.697$ \\  
Betweenness  & $-20.40$ & $0.062$ & $-26.22$ & $0.043^{*}$\\ 
Closeness & \hspace{1mm}$113.36$ & $0.037^{*}$ & \hspace{2mm}\,$94.82$ & $0.032^{*}$ \\ 
ClosenessINS & \hspace{1mm}$119.63$ & $0.035^{*}$ & $\hspace{1mm}100.09$ & $0.030^{*}$ \\ \hline 
\end{tabular}
\label{tab:centr-all}
\end{table}

As shown in Table \ref{tab:centr-all}, we found statistically significant positive correlations with a degree, indegree and closeness for persistence in physics. For MI we found statistically significant correlations only for measures considering the entire social network, that is betweenness and closeness, and no statistically significant correlations for measures aimed at the students ego network. 

\begin{table}[t]
\renewcommand{\arraystretch}{1.2}
\renewcommand{\tabcolsep}{6pt}
\caption{Summary of the likelihood ratio test performed for the multivariate logistics regression with a student's final grade considered as additional predictors of the persistence when compared to the simple models.}
\begin{tabular}{p{0.3cm}p{3.6cm}p{0.5cm}p{0.8cm}p{1.2cm}}
 \hline
\multicolumn{2}{c}{Model $\sim$ Centrality + Grade} & df &  $\hspace{2mm}\chi^2$ &  p-value  \\ \hline
\parbox[t]{2mm}{\multirow{4}{*}{\rotatebox[origin=c]{90}{In physics}}} & \hspace{7mm}Total degree &2 &  25.75 & $<10^{-6}$ \\
 &  \hspace{7mm}In degree &1 & 25.83 & $<10^{-6}$\\
  &  \hspace{7mm}Closeness &1 &26.63 &$<10^{-6}$ \\
&  \hspace{7mm}ClosenessINS&1&26.58&$<10^{-6}$\\ \hline
\parbox[t]{2mm}{\multirow{3}{*}{\rotatebox[origin=c]{90}{In MI}}}  & \hspace{7mm}Betweenness &$1$   & $15.28$ & $<10^{-4}$\\
 &  \hspace{7mm}Closeness &1 &11.17 & $<10^{-3}$\\
 &  \hspace{7mm}ClosenessINS  &1&11.14&$<10^{-3}$\\ \hline
\label{tab:nested-models}
\end{tabular}
\end{table}

To determine whether our univariate models can be improved we considered nested multivariate models for all the statistically significant centrality measures, with a student's gender, ethnicity, academic plan (declared major) and a final grade considered as additional predictors of the persistence. We found that only for the grade made a statistically significant difference in the model fits. Table \ref{tab:nested-models} summarizes the results of the logistic regression for both in physics and in MI cases. However, when we compared the fit of the multivariate models to the fit of the models reduced to a grade as a sole predicting variable, we found significantly better fit only for the full betweenness model ($\chi^2(1)=7.89$, $p=0.005$). The variance inflation factors indicates the lack of collinearity among betweenness and grade ($VIF = 1.03$).

Finally, to optimize the correlation and to determine the predictability threshold for centralities we used the mutual information. Table~\ref{tab:mutual-inf} shows the threshold values for each centrality measure and its significance level.

\section{Discussion}

\begin{table}[b]
\renewcommand{\arraystretch}{1.2}
\renewcommand{\tabcolsep}{4pt}
\caption{The threshold value ($\theta$) for each centrality measure as determined by maximization of the mutual information and its significance level measured by the chi-square test.}
\begin{tabular}{p{0.3cm}p{2.6cm}p{1cm}p{0.5cm}p{0.8cm}p{1.2cm}} \hline
\multicolumn{2}{c}{Centrality} & \hspace{3mm}$\theta$ & df &  $\hspace{2mm}\chi^2$ &  p-value  \\ \hline
\parbox[t]{2mm}{\multirow{4}{*}{\rotatebox[origin=c]{90}{In physics}}} & \hspace{3mm}Total degree & \hspace{3mm}1& 1 & 11.05 & $<10^{-3}$ \\
& \hspace{3mm}In degree &\hspace{3mm}1 & 1 & \hspace{2mm}7.37 & $\hspace{2mm}0.007$ \\
& \hspace{3mm}Closeness & 0.013 & 1 &\hspace{2mm}8.62 &$\hspace{2mm}0.003$ \\
&  \hspace{3mm}ClosenessINS & 0.012 & 1& 11.61 & $<10^{-3}$\\ \hline
\parbox[t]{2mm}{\multirow{2}{*}{\rotatebox[origin=c]{90}{In MI}}} & \hspace{3mm}Closeness & 0.022 & 1 & \hspace{2mm}4.53 & $0.033$\\
& \hspace{3mm}ClosenessINS  & 0.021 & 1 & \hspace{2mm}4.53 & $0.033$\\ \hline
\label{tab:mutual-inf}
\end{tabular}
\end{table}

We find that students with higher certain centrality measures at the end of the first semester of MI are in fact more likely to enroll in a second semester of physics. For the MI sequence, we found that students with low closeness seem to be more likely to enroll in a second semester of MI while (in/out/total)degree has no affect on their decision. On the other hand, students with high betweenness score tend to either switch to traditional curriculum or to leave physics altogether. Moreover, higher grades strengthen this negative correlation, that is students with higher final grades and high betweenness are the most likely to leave MI but remain in physics.

To explain this discrepancy one needs to understand what these two measures mean in practice. Closeness can be thought of as strong embeddedness within the entire network. Students with low closeness scores are close to all the other students in the network and thus they have an easy access to information from many sources. They are also -- by sheer nature of this measure -- connected to many students. This can help them appreciate all the benefits of having a strong network of connection within a classroom. Betweenness, on the other hand, depends mainly on the position within the network. In practice, in order to have high betweenness it suffices to be connecting clusters otherwise separate. Thus, students with high betweenness score are not necessarily connected to many other students.  

For physics in general we found a statistically significant positive correlation between (in)degree and closeness. However, due to a small sample of students who took a non-MI physics this finding requires further study.

It should be kept in mind that a centrality which is appropriate for one category will often ``get it wrong'' when applied to a different category. More importantly, while centralities identify the most important vertices in a given network, this ranking cannot be generalized to the remaining vertices with lower scores - centrality does not indicate the relative importance of all vertices.

While the correlation with increased persistence is an ongoing study, these findings suggest that student social integration influences persistence.


\begin{acknowledgments}
We would like to thank Geoff Potvin for facilitating data collection, Eric Williams for helpful discussions and Anita D\k{a}browska for feedback and recommendations on statistical analyses. Supported by NSF PHY 1344247.
\end{acknowledgments}


\bibliographystyle{plain}

\begin{thebibliography}{13}
\expandafter\ifx\csname natexlab\endcsname\relax\def\natexlab#1{#1}\fi
\providecommand{\enquote}[1]{``#1''}
\expandafter\ifx\csname url\endcsname\relax
  \def\url#1{\texttt{#1}}\fi
\expandafter\ifx\csname urlprefix\endcsname\relax\def\urlprefix{URL }\fi
\providecommand{\eprint}[2][]{\url{#2}}

\bibitem[Tinto (1975)]{Tinto75-DHE}
V.~Tinto, Rev. Educ. Res. \textbf{45}, 1 (1975).

\bibitem[Tinto (1997)]{Tinto97-CAC}
V.~Tinto, J. High. Educ. \textbf{68}, 6 (1997).

\bibitem[Wasserman et~al(1994)]{Wasserman-SNA}
S.~Wasserman and K.~Faust,  \emph{Social Network Analysis: Methods and Applications}, (Cambridge Univ. Press, 1994).

\bibitem[Scott et~al(2011)]{Scott-SNA}
J.~Scott and P.~J.~Carrington, eds.  \emph{The SAGE Handbook of Social Network Analysis}, (SAGE Publications Ltd, 2011).

\bibitem[Brewe et~al.(2012)]{Brewe12-SNA}
E.~Brewe L.~Kramer and V.~Sawtelle, , Phys. Rev. ST PER \textbf{8}, 010101 (2012).

\bibitem[Forsman et~al.(2014)]{Forsman14-CSN}
J.~Forsman, R.~Moll and C.~Linder, Phys. Rev. ST. PER \textbf{10}, 020122 (2014).

\bibitem{Note1}
In this contexts, \emph{important} can mean the ability to transfer information across the network or it can be understood as involvement in the cohesiveness of the network.


\bibitem{Note3}
The mutual information is an information theoretic measure of dependency between two random variables.

\bibitem[R(2015)]{R}
R Core Team (2015). R: A language and environment for statistical computing. R Foundation for Statistical Computing, Vienna, Austria.



\end{thebibliography}

\IfFileExists{\jobname.bbl}{}
 {\typeout{}
  \typeout{******************************************}
  \typeout{** Please run "bibtex \jobname" to optain}
  \typeout{** the bibliography and then re-run LaTeX}
  \typeout{** twice to fix the references!}
  \typeout{******************************************}
  \typeout{}
 }

\end{document}